\documentclass[a4paper,10pt]{article}
\usepackage{graphicx}
\usepackage{comment}
\usepackage{amssymb}
\usepackage{amsmath}
\usepackage{nicefrac}
\usepackage{graphicx}
\usepackage{dcolumn}
\usepackage{bm}
\usepackage{comment}
\usepackage{multirow}
\usepackage{cite}

\begin{document}

\newcommand{\bin}[2]{\left(\begin{array}{c}\!#1\!\\\!#2\!\end{array}\right)}
\newcommand{\threej}[6]{\left(\begin{array}{ccc}#1 & #2 & #3 \\ #4 & #5 & #6 \end{array}\right)}
\newcommand{\sixj}[6]{\left\{\begin{array}{ccc}#1 & #2 & #3 \\ #4 & #5 & #6 \end{array}\right\}}
\newcommand{\regge}[9]{\left[\begin{array}{ccc}#1 & #2 & #3 \\ #4 & #5 & #6 \\ #7 & #8 & #9 \end{array}\right]}
\newcommand{\La}[6]{\left[\begin{array}{ccc}#1 & #2 & #3 \\ #4 & #5 & #6 \end{array}\right]}
\newcommand{\hj}{\hat{J}}
\newcommand{\hux}{\hat{J}_{1x}}
\newcommand{\hdx}{\hat{J}_{2x}}
\newcommand{\huy}{\hat{J}_{1y}}
\newcommand{\hdy}{\hat{J}_{2y}}
\newcommand{\huz}{\hat{J}_{1z}}
\newcommand{\hdz}{\hat{J}_{2z}}
\newcommand{\hup}{\hat{J}_1^+}
\newcommand{\hum}{\hat{J}_1^-}
\newcommand{\hdp}{\hat{J}_2^+}
\newcommand{\hdm}{\hat{J}_2^-}

\huge

\begin{center}
Simple electron-impact excitation cross-sections including plasma density effects
\end{center}

\vspace{0.5cm}

\large

\begin{center}
Jean-Christophe Pain$^{a,b,}$\footnote{jean-christophe.pain@cea.fr} and Djamel Benredjem$^c$
\end{center}

\normalsize

\begin{center}
\it $^a$CEA, DAM, DIF, F-91297 Arpajon, France\\
\it $^b$Universit\'e Paris-Saclay, CEA, Laboratoire Mati\`ere en Conditions Extr\^emes,\\
\it 91680 Bruy\`eres-le-Ch\^atel, France\\
\it $^c$ Laboratoire Aim\'e Cotton, Universit\'e Paris-Saclay, Orsay, France
\end{center}

\vspace{0.5cm}

\begin{abstract}
The modeling of non-local-thermodynamic-equilibrium plasmas is crucial for many aspects of high-energy-density physics. It often requires collisional-radiative models coupled with radiative-hydrodynamics simulations. Therefore, there is a strong need for fast and as accurate as possible calculations of the cross-sections and rates of the different collisional and radiative processes. We present an analytical approach for the computation of the electron-impact excitation (EIE) cross-sections in the Plane Wave Born (PWB) approximation. The formalism relies on the screened hydrogenic model. The EIE cross-section is expressed in terms of integrals, involving spherical Bessel functions, which can be calculated analytically. In order to remedy the fact that the PWB approximation is not correct at low energy (near threshold), we consider different correcting factors (Elwert-Sommerfeld, Cowan-Robb, Kilcrease-Brookes). We also investigate the role of plasma density effects such as Coulomb screening and quantum degeneracy on the EIE rate. This requires to integrate the collision strength multiplied by the Fermi-Dirac Distribution and the Pauli blocking factor. We show that, using an analytical fit often used in collisional-radiative models, the EIE rate can be calculated accurately without any numerical integration, and compare our expression with a correction factor presented in a recent work.
\end{abstract}

\section{Introduction}\label{sec1}

Interpretation of spectroscopic measurements and simulations of kinetic and transport processes in non-local-thermodynamic-equilibrium (NLTE) plasmas requires knowledge of many electron-impact excitation (EIE) cross-sections for atoms and ions. This is the case, for instance, for integrated simulations of hohlraums in inertial confinement fusion, diagnosis of plasma X-ray sources, estimation of radiative power losses in magnetic-confinement-fusion devices or photoionized plasmas in astrophysics. The modeling of atomic physics in plasmas out of equilibrium depends closely on the radiation field and radiation transport and is generally coupled to the hydrodynamic motion of matter, so that NLTE physics must be used in integrated radiation-hydrodynamics simulations. This requires fast but accurate calculations of cross-sections and rates for the processes involved in collisional-radiative models (see for instance \cite{CHUNG05,SCOTT10,BENITA15,VICHEV19}). As an example, we mention the code ATMED CR, which includes relativistic $n\ell j$ splitting as well as non zero $\Delta n$ and elastic $\Delta n=0$ collisions with plasma electrons \cite{benita,benita17,benita19}. Those codes are helpful for checking orders of magnitudes and making comparisons between numerical values of rates and other data relevant for the modeling of NLTE plasmas.

Since the measurements of EIE cross-sections in dense plasmas are definitely scarce, it is difficult to provide prescriptions concerning screening charges, near-threshold corrections or energy-level shifts. Bearing in mind this fact, the rates of ATMED CR are really good. In addition, inertial confinement fusion seeks to compress material at low temperatures and high densities. Under these conditions, free-electron degeneracy can have a significant effect on plasma ionization, emission and absorption. 

Two classes of methods are commonly used in the calculation of EIE cross-sections. The first is based on a set of close-coupling (CC) equations, which takes into account the coupling of various excitation channels \cite{SEATON76}. In these approaches, resonances can be included in a natural way by taking into account the coupling to closed channels. Several implementations exist. The most widely used is the R-matrix code developed by a group at the Queens University of Belfast \cite{BERRINGTON95}. The second class of methods is based on the first-order Born approximation, which assumes independent excitation channels. The coupling to closed channels, which results in resonances, may be included with the help of perturbation methods \cite{EISSNER72}. Different variants exist according to the treatment of the continuum wave functions. The plane wave Born (PWB) approximation uses unperturbed plane waves for the free orbitals. The Coulomb wave Born (CWB) approximation \cite{SEATON62} takes into account the distortion of the continuum wave-functions due to a pure Coulomb potential. The most accurate of this class is the Distorted Wave  Born (DWB) approximation, in which the free orbitals are calculated in a more realistic potential taking into account the electronic structure of the target ion. The majority of the computer programs implement the DW approximation \cite{PAIN19b} , since it yields significantly better results than the PW and CW methods with minimal increase in complexity. Many DW codes are in use today such as the non-relativistic code from University College London \cite{EISSNER98}, the HULLAC package \cite{BAR88}, the code by Zhang \textit{et al.} \cite{ZHANG89}, that of Chen \cite{CHEN96} or the FAC code \cite{GU08}, just to name a few. 

In this work, we present a simple and rather accurate approach for the computation of the EIE cross-section in the PWB approximation. The formalism relies on the screened hydrogenic model. The EIE cross-section is expressed in terms of integrals involving spherical Bessel functions which can be calculated analytically. In order to remedy the fact that the PWB approximation is not correct at low energy (near threshold), we also compare different correcting factors (Elwert-Sommerfeld \cite{SOMMERFELD53,BIEDENHARN56,BETHE57}, Cowan-Robb \cite{COWAN81}, Kilcrease-Brookes \cite{KILCREASE13}).

One of the most significant ``standard'' contributions to the shift of H-like spectral lines is caused by quenching non-zero $\Delta n$ \cite{GRIEM83} and elastic $\Delta n=0$ \cite{BOERCKER84} collisions with plasma electrons, the so-called electronic shift \cite{GRIEM88,BENREDJEM90}. There is also a so-called Plasma Polarization Shift (PPS), which plays an important role in explaining the observed shifts of the high-$n$ H-like spectral lines \cite{IGLESIAS95,RENNER98}. Physically, the PPS is caused by the redistribution of plasma electrons due to interaction with the radiating ion. When only plasma electrons inside the orbit of the bound electron were taken into account, the resulting PPS was expected to be towards the blue \cite{BERG62}. If the free electrons outside the bound-electron orbit are also taken into account, the resulting PPS is towards the red. The theoretical results for red PPS by different authors differ by a factor of two \cite{OKS17}. The ionization potential depression (IPD) in a dense plasma is somehow an average quantity characterizing the global effect of the charged particles $-$perturbers$-$ on a given ion. Quantum properties, such as the ionization potential are modified due to the interactions of the valence electron with the perturbers. Two models, namely the Stewart-Pyatt (SP) \cite{STEWART66} and Ecker-Kr\"oll (EK) \cite{ECKER63} models have been widely used during the past decades to estimate the IPD. A few years ago, their validity has been discussed in the framework of two experiments, one using an X-ray free-electron laser \cite{CIRICOSTA12} and the other one using a high-power optical laser \cite{HOARTY13} to create the dense plasma. It appeared that neither the SP model nor the EK model were able to explain both experiments. This has initiated a renewed interest for the problem of the IPD in dense plasmas (see for example Refs. \cite{DEBYE23,ZAGHLOUL09,VINKO14,IGLESIAS14,SON14,CALISTI15b}). 

We present a model for the computation of EIE cross-sections which has the advantage of being almost completely analytical (this was our ``requirements specification''). It can be useful for fast ``on-line'' NLTE calculations in radiative-hydrodynamics simulations for instance. The source code is available upon request. It relies on the screened hydrogenic approximation and enables one to take into account the impact of some plasma density effects on the EIE cross-section, such as electronic shifts or degeneracy effects. In particular, we suggest to model the effect of electrons (which leads to a blue shift) by extending the Li-Rosmej formula \cite{ROSMEJ11,LI12,IGLESIAS19a,PAIN19}. The energy shift due to ions (which is in fact a red shift) will not be considered here. It can be modeled using an approach similar to the one published recently by Iglesias \cite{IGLESIAS19a}. The range of validity of the model is difficult to establish; owing to the  assumptions and approximations made, our approach will give reliable results when the screened hydrogenic approximation is reliable. It is therefore strongly dependent on the quality of the mean charge and subsequently the screening constants. Since it is not simple to determine unambiguously which set of screening constants is the best, we have investigated the impact of different sets (see  Refs. \cite{MAYER47,FAUSSURIER97,SMITH11}). The two latter depend on principal $n$ and orbital $\ell$ quantum numbers. Note that relativistic screening constants (depending on $n$, $\ell$ and $j$) were derived by several authors (see for instance Ref. \cite{mendoza11}), but in the present case we restrict ourselves to the non-relativistic hydrogenic approximation. It is definitely true that Mayer's constants \cite{MAYER47} depend only on $n$, but they are the ones for which, in that particular case (transition in Li-like carbon), we got the best agreement with the quantum-mechanical calculations performed using Cowan's code. This is definitely surprising, but cannot be generalized to other cases, of course, since it might be due to compensation of errors due to the approximations of our model.

In section \ref{sec2}, we present our model for the EIE cross-section calculation. In section \ref{sec3}, comparisons with DW calculations are performed and discussed. A special care is given to the behaviour near threshold, where several correction factors are compared. In section \ref{sec4} the impact of energy shifts due to ions and electrons is studied using analytical formulas.

\section{Calculation of the electron-impact cross-section}\label{sec2}

\subsection{Plane Wave Born approximation}\label{subsec21}

\noindent Let us denote respectively $a$ and $b$ the initial and final states of the transition induced by the electron impact. In the PWB approximation, the EIE cross-section reads (in atomic units):

\begin{equation}\label{XSection}
\sigma(\epsilon)=\pi a_0^2\frac{8}{g_a}\sum_{M,M'}\int_{k_{\mathrm{min}}}^{k_{\mathrm{max}}}\frac{|\langle\gamma JM|\sum_j e^{i\textbf{k}\cdot\textbf{r}_j}|\gamma'J'M'\rangle|^2}{k^3}\mathrm{d}k,
\end{equation}
where $k=\sqrt{2\epsilon}$ is the wavevector and $\epsilon$ the energy of the incident electron. $k_{\mathrm{min}}=\sqrt{2\epsilon}-\sqrt{2(\epsilon-\Delta\epsilon)}$ and $k_{\mathrm{max}}=\sqrt{2\epsilon}+\sqrt{2(\epsilon-\Delta\epsilon)}$, $\Delta\epsilon$ representing the excitation energy. The initial and final states of the ion are respectively $|\gamma JM\rangle$ and $|\gamma'J'M'\rangle$ and $\textbf{r}_j$ is the position of the $j^{th}$ electron. $J$ is the total atomic angular momentum, $M$ its projection on the $z-$axis and $\gamma$ represents all the additional required quantum numbers in order to define the state in an unique way. The coupling of all the quantum numbers included in $\gamma$ leads to $J$. $g_a$ represents the degeneracy of the initial state. In the present study, we restrict ourselves to dipole-allowed transitions, but the approach can be generalized to non-dipole transitions. This is an advantage over the Van Regemorter formula \cite{VANREGEMORTER62,FISHER96}. The cross-section can be written as

\begin{equation}\label{regemo}
\sigma(\epsilon)=\frac{\pi a_0^2}{g_a\epsilon}~\Omega(\epsilon),
\end{equation}

where the collision strength reads

\begin{equation*}
\Omega(\epsilon)=\frac{8}{\Delta\epsilon}\int_{k_{\mathrm{min}}}^{k_{\mathrm{max}}}gf(k)d(\ln k),
\end{equation*}

\noindent the quantity $gf(k)$ being the generalized oscillator strength. In the hydrogenic approximation, $a\equiv n_a\ell_a$, $b\equiv n_b\ell_b$ and the matrix element in Eq. (\ref{XSection}) can be written $\langle n_a\ell_a m_a|e^{i\textbf{k}.\textbf{r}}|n_b\ell_bm_b\rangle$ and calculated analytically, as will be shown in the next section.

\subsection{Screened hydrogenic model}\label{subsec22}

Assuming $\bf{k}$=$k\bf{e}_z$, we have :

\begin{equation*}
e^{i\textbf{k}\cdot\textbf{r}}=e^{ikz}=e^{ikr\cos\theta}=\sum_{t=0}^{\infty}i^{t}(2t+1)^{1/2}(4\pi)^{1/2}j_{t}(kr)Y_{t}^0(\theta)
\end{equation*}

\noindent and therefore

\begin{eqnarray}\label{sum}
\langle n_a\ell_a m_a|e^{i\textbf{k}\cdot\textbf{r}}|n_b\ell_bm_b\rangle&=&\sum_{\ell=0}^{\infty}i^{\ell}(2\ell+1)^{1/2}(4\pi)^{1/2}\mathcal{M}_r(n_a,\ell_a,t,k,n_b,\ell_b)\mathcal{N}_a(\ell_a,m_a,t,0,\ell_b,m_b),
\end{eqnarray}

\noindent where

\begin{equation*}
\mathcal{M}_r(n_a,\ell_a,t,k,n_b,\ell_b)=\int_0^{\infty}R_{n_a\ell_a}(r)j_{t}(kr)R_{n_b\ell_b}(r)r^2\mathrm{d}r
\end{equation*}

\noindent and

\begin{equation*}
\mathcal{N}_a(\ell_a,m_a,t,0,\ell_b,m_b)=\int Y_{\ell_a}^{m_a*}Y_{t}^0Y_{\ell_b}^{m_b}d^2\Omega,
\end{equation*}

where $Y_{\ell}^{m}(\theta,\phi)$ are normalized spherical harmonics and the $R_{n\ell}(r)$ radial hydrogenic wave functions. $d^2\Omega=\sin\theta d\theta d\phi$ is the infinitesimal solid angle. Angular integrals of the type $\mathcal{N}_a$ are well-known (``Gaunt coefficients'') and analytical expressions may be found in the current literature. In particular, the Gaunt coefficients can be expressed in terms of $3j$ coefficients:

\begin{equation*}
\int Y_{\ell_1}^{m_1}Y_{\ell_2}^{m_2} Y_{\ell_3}^{m_3}d^2\Omega=\sqrt{\frac{(2\ell_1+1)(2\ell_2+1)(2\ell_3+1)}{4\pi}}\threej{\ell_1}{\ell_2}{\ell_3}{m_1}{m_2}{m_3}\threej{\ell_1}{\ell_2}{\ell_3}{0}{0}{0}.
\end{equation*}

\noindent Since

\begin{equation*}
\int Y_{\ell_1}^{m_1}Y_{\ell_2}^{m_2}Y_{\ell_3}^{m_3}d^\Omega=(-1)^{m_2}\int Y_{\ell_1}^{m_1}Y_{\ell_2}^{-m_2*} Y_{\ell_3}^{m_3}d^2\Omega,
\end{equation*}

\noindent we get

\begin{eqnarray*}
\int Y_{\ell_1}^{m_1}Y_{\ell_2}^{m_2*} Y_{\ell_3}^{m_3}d^2\Omega&=&(-1)^{m_2}\sqrt{\frac{(2\ell_1+1)(2\ell_2+1)(2\ell_3+1)}{4\pi}}\threej{\ell_1}{\ell_2}{\ell_3}{m_1}{m_2}{m_3}\threej{\ell_1}{\ell_2}{\ell_3}{0}{0}{0}.\nonumber\\
&=&\sqrt{\frac{(2\ell_1+1)(2\ell_3+1)}{4\pi (2\ell_2+1)}}\langle \ell_10|\ell_20\ell_30\rangle\langle\ell_1m_1|\ell_2m_2\ell_3m_3\rangle,
\end{eqnarray*}

\noindent where $\langle\ell_1m_1|\ell_2m_2\ell_3m_3\rangle$ is the usual Clebsch-Gordan coefficient, and then

\begin{eqnarray*}
\mathcal{N}_a(\ell_a,m_a,t,0,\ell_b,m_b)&=&\int Y_{\ell_a}^{m_a*}Y_{t}^0Y_{\ell_b}^{m_b}d^2\Omega=(-1)^{m_a}\sqrt{\frac{(2t+1)(2\ell_a+1)(2\ell_b+1)}{4\pi}}\nonumber\\
& &\times\threej{t}{\ell_a}{\ell_b}{0}{m_a}{m_b}\threej{t}{\ell_a}{\ell_b}{0}{0}{0}\nonumber\\
&=&\sqrt{\frac{(2\ell_a+1)(2\ell_b+1)}{4\pi (2t+1)}}\langle \ell_a0|t0\ell_b0\rangle\langle\ell_am_a|t0\ell_bm_b\rangle.
\end{eqnarray*}

\noindent The latter integral vanishes unless $|\ell_a-\ell_b|\leq t\leq\ell_a+\ell_b$ and therefore the sum in Eq. (\ref{sum}) is not infinite. Thus, one can write

\begin{equation*}
gf(k)=\frac{\Delta\epsilon}{k^2}(2\ell_a+1)(2\ell_b+1)\sum_t(2t+1)\left[\threej{\ell_a}{t}{\ell_b}{0}{0}{0}\int_0^{\infty}P_a(r)j_t(kr)P_b(r)\mathrm{d}r\right]^2,
\end{equation*}

\noindent with $|\ell_a-\ell_b|\leq t\leq\ell_a+\ell_b$, $\mathrm{mod}(\ell_a+t+\ell_b,2)=0$, and $P_a(r)=rR_{n_a\ell_a}(r)$. The radial wave-functions are defined as

\begin{equation}\label{rn}
R_{n\ell}(r)=\sqrt{\frac{Z_{n\ell}(n-\ell-1)!}{n^2(n+\ell)!}}\left(\frac{2Z_{n\ell}}{n}\right)^{\ell+1}r^{\ell}\exp\left[-\frac{rZ_{n\ell}}{n}\right]L_{n-\ell-1}^{2\ell+1}\left(\frac{2rZ_{n\ell}}{n}\right),
\end{equation}

\noindent where $Z_{n\ell}$ is the screened nuclear charge seen by an electron in the $n\ell$ subshell, and $L_{n-\ell-1}^{2\ell+1}(x)$ a generalized Laguerre polynomial. The energy of state $i$ is, in atomic units

\begin{equation*}
\epsilon_i=-\frac{Z_{n_i\ell_i}^2}{2n_i^2}.
\end{equation*}

\subsection{Analytical form of the generalized oscillator strength}\label{subsec23}

The generalized Laguerre polynomial can be expanded as 

\begin{equation}\label{explag}
L_s^t(x)=\sum_{j=0}^s\bin{t+s}{s-j}\frac{(-x)^j}{j!},
\end{equation}

\noindent where $\bin{m}{n}=\frac{m!}{n!(m-n)!}$ is the usual binomial coefficient. Eq. (\ref{explag}) yields

\begin{eqnarray*}
L_{n_a-\ell_a-1}^{2\ell_a+1}\left(\frac{2rZ_a}{n_a}\right)L_{n_b-\ell_b-1}^{2\ell_b+1}\left(\frac{2rZ_b}{n_b}\right)&=&\sum_{j=0}^{n_a-\ell_a-1}\sum_{u=0}^{n_b-\ell_b-1}\bin{n_a+\ell_a}{n_a-\ell_a-1-j}\bin{n_b+\ell_b}{n_b-\ell_b-1-u}\nonumber\\
& &\times\frac{(-1)^{j+u}}{j!u!}\left(\frac{2Z_a}{n_a}\right)^j\left(\frac{2Z_b}{n_b}\right)^ur^{j+u}.\nonumber\\
& &
\end{eqnarray*}

\noindent The latter formula enables one to obtain the following expression for the generalized oscillator strength:

\begin{eqnarray*}
gf(k)&=&\frac{\Delta \epsilon}{k^2}\left(2\ell_a+1\right)\left(2\ell_b+1\right)\frac{Z_aZ_b}{n_a^2n_b^2}\frac{\left(n_a-\ell_a-1\right)!\left(n_b-\ell_b-1\right)!}{\left(n_a+\ell_a\right)!\left(n_b+\ell_b\right)!}\nonumber\\
& &\times\left(\frac{2Z_a}{n_a}\right)^{2\ell_a+2}\left(\frac{2Z_b}{n_b}\right)^{2\ell_b+2}\sum_t(2t+1)\threej{\ell_a}{t}{\ell_b}{0}{0}{0}^2\left\{\sum_{j=0}^{n_a-\ell_a-1}\sum_{u=0}^{n_b-\ell_b-1}\bin{n_a+\ell_a}{n_a-\ell_a-1-j}\right.\nonumber\\
& &\times\bin{n_b+\ell_b}{n_b-\ell_b-1-u}\frac{(-1)^{j+u}}{j!u!}\left(\frac{2Z_a}{n_a}\right)^j\left(\frac{2Z_b}{n_b}\right)^u\nonumber\\
& &\times\left. \mathcal{I}\left(t,j+u+\ell_a+\ell_b+2,\frac{Z_a}{n_a}+\frac{Z_b}{n_b},k\right)\vphantom{\sum_{j=0}^{n_a-\ell_a-1}\sum_{u=0}^{n_b-\ell_b-1}\bin{n_a+\ell_a}{n_a-\ell_a-1-j}}\right\}^2,\nonumber\\
& &
\end{eqnarray*}

\noindent $\mathcal{I}$ being an integral of the type

\begin{equation*}
\mathcal{I}(\alpha,\beta,c,d)=\int_0^{\infty}e^{-cr}j_{\alpha}(dr)r^{\beta}\mathrm{d}r,
\end{equation*}

\noindent where $c$ and $d$ are real and $\beta$ is an integer such that $\beta>\alpha$. $\mathcal{I}$ can be put in the form

\begin{eqnarray}\label{fin}
\mathcal{I}(\alpha,\beta,c,d)&=&\frac{1}{c^{\beta+1}}\frac{(2w)^{\alpha}}{\left(1+w^2\right)^{\beta}}\left\{\frac{1+\sqrt{1+w^2}}{2}\right\}^{\beta-\alpha-1}\frac{\alpha!(\alpha+\beta)!}{(2\alpha+1)!}\nonumber\\
& &\times\sum_{i=0}^{\beta-\alpha-1}\frac{\bin{\beta-\alpha-1}{i}\bin{\alpha+i}{i}\bin{2\beta-1}{2i}}{\bin{\beta-1}{i}\bin{2\alpha+2i+1}{2i}}\left[\frac{1-\sqrt{1+w^2}}{1+\sqrt{1+w^2}}\right]^i,
\end{eqnarray}

\noindent with $w=d/c$. One has also

\begin{equation}
\mathcal{I}(\alpha,\beta,c,d)=\frac{w^{\alpha}\sqrt{\pi}}{2^{1+\alpha}c^{1+\beta}}\,\frac{(\alpha+\beta)!}{\Gamma(\alpha+3/2)}\,_2F_1\left[\begin{array}{c}
    \frac{\alpha+\beta+1}{2},\frac{\alpha+\beta}{2}+1\nonumber\\
    \alpha+\frac{3}{2}\end{array};-w^2
    \right],    
\end{equation}

\noindent where $_2F_1$ is the usual Gauss hypergeometric function. Expression (\ref{fin}) is simpler than the formulation proposed by Upcraft \cite{UPCRAFT10}. In the latter work, the author expressed the spherical Bessel functions in terms of their sine and cosine terms which implied to handle many quantities of the type

\begin{equation*}
\int_0^{\infty}e^{-cr}\sin(dr)r^{\beta}\mathrm{d}r
\end{equation*}

\begin{figure*}
\vspace{1cm}
\begin{center}
\includegraphics[width=10cm]{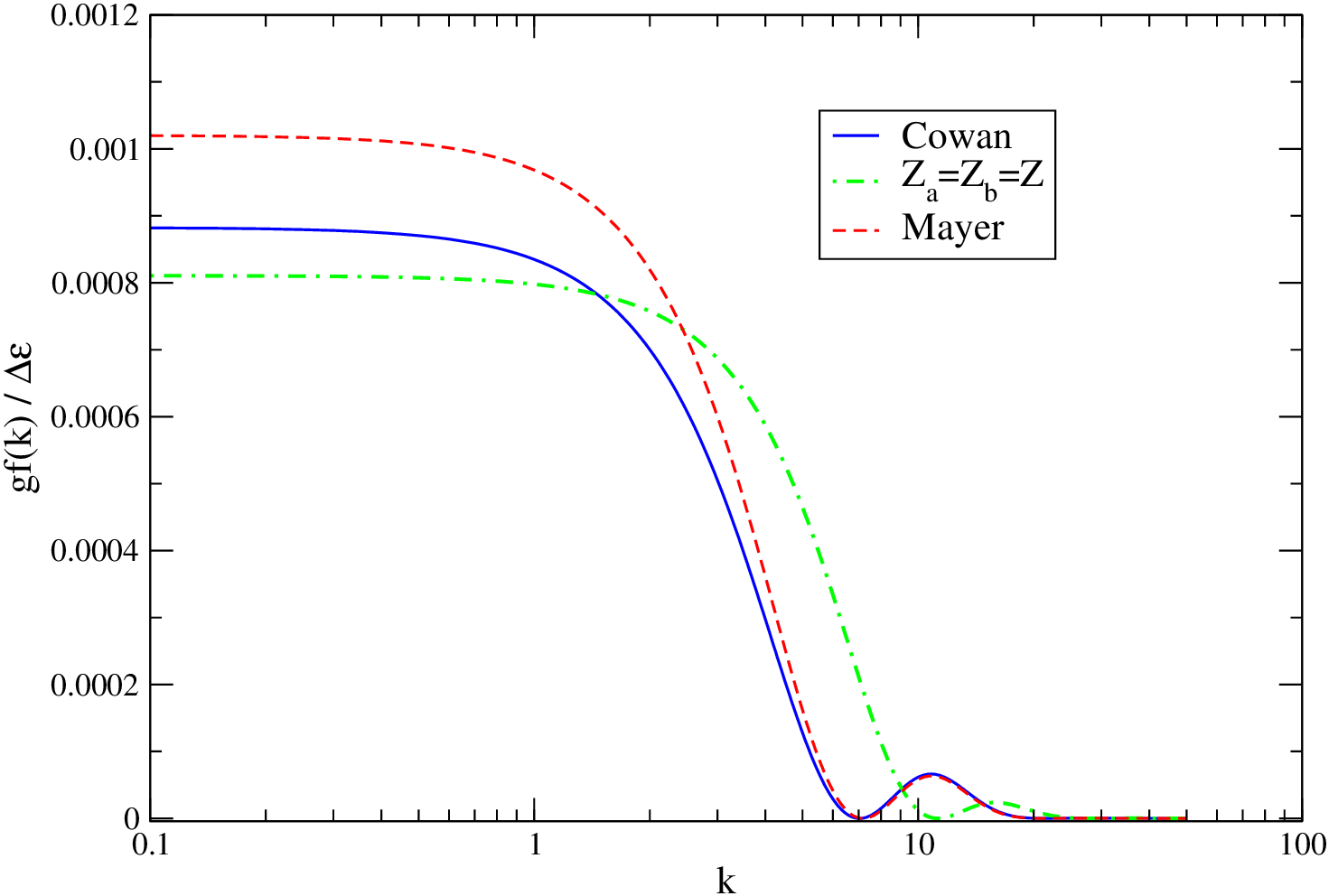}
\end{center}
\caption{(Color online) $gf(k)/\Delta\epsilon$ (at. units) for transition $2s\rightarrow 4p$ in [Ne] Fe XVII. Blue curve: Cowan's code computation \cite{COWAN81}, green dot-dashed curve: unscreened case ($Z_a=Z_b=Z$), red dashed curve: use of Mayer's screening constants \cite{MAYER47}.}\label{upcraft_fig4}
\vspace{1cm}
\end{figure*}

\noindent which is much more tedious than using the compact expression (\ref{fin}). Figure \ref{upcraft_fig4} displays $gf(k)/\Delta\epsilon$ for the transition $2s\rightarrow 4p$ in [Ne] Fe XVII. We compare three calculations: one represents the present model without screening constants ($Z_a=Z_b=Z$, green dot-dashed curve), the second one represents the present model using the screening constants published by Mayer \cite{MAYER47} (green dashed curve) and the third one is a quantum-mechanical calculation (but still in the PWB approximation) performed with Cowan's code (in blue) considered here as the reference. In this specific example, the asymptote for small $k$ is better reproduced by the choice $Z=Z_a=Z_b$ (\textit{i.e.} no screening constants), but the intermediate region (for $k$ between 3 and 7) as well as the bump around $k$=10 reveal that the screening constant improves the agreement with Cowan's code.

\section{Corrections near threshold}\label{sec3}

We now address the issue of the non-validity of the PWB approach near threshold. Since our expression of the collision strength is obtained within the framework of the screened hydrogenic approximation, we get configuration-averaged collision strengths from the one-electron form by:

\begin{equation*}
\Omega_{i\rightarrow f}(\epsilon)=\frac{2q_{i}(g_f-q_f)}{g_{i}g_f}\Omega(\epsilon),
\end{equation*}

\noindent $q_i$ and $g_i$ being respectively the population and the degeneracy of the active subshell $i$. The FAC code \cite{GU08} gives the collision strength between all the fine-structure levels of two configurations. In order to sum the fine-structure collision strengths between the $\gamma J$ levels of the lower configuration $C$ to the $\gamma'J'$ levels of the upper configuration $C'$ into the configuration-averaged value, we have

\begin{equation*}
\Omega_{i\rightarrow f}(\epsilon)=\frac{1}{G_C}\sum_{\gamma J\in C}\sum_{\gamma' J'\in C'}g_{\gamma J}\Omega_{\gamma J\rightarrow\gamma'J'}(\epsilon),
\end{equation*}

\noindent where $g_{\gamma J}$ is the degeneracy of the $\gamma J$ level and $G_C$ the degeneracy of the lower configuration, \textit{i.e.}

\begin{equation*}
G_C=\prod_{i\in C}\bin{g_i}{q_i},
\end{equation*}

\noindent the product being taken over all occupied subshells of the lower configuration.

The so-called Cowan-Robb correction to the PWB cross-section (see Ref. \cite{COWAN81}, pp. 568-569) consists in replacing $\Omega(\epsilon)$ by $\Omega_m(\epsilon)$ which is defined as 

\begin{equation*}
\Omega_m(\epsilon)=\Omega\left(\frac{\epsilon}{\Delta\epsilon}+\frac{3}{1+\epsilon/\Delta \epsilon}\right),
\end{equation*}

\noindent in order to extrapolate the collision strength $\Omega(E)$ from above threshold down to the threshold. In Ref. \cite{COWAN81}, Cowan points out that this prescription is tentative, valid only for spin-allowed transitions and was obtained by comparison with more accurate results relying on close-coupling and DW approaches.

Kim proposed a re-scaling of the PWB cross-sections of neutral atoms \cite{KIM02a}, and of CWB cross-sections for singly charged ions. In the latter case, it consists in multiplying the CWB cross-section without exchange by the factor $\epsilon/(\epsilon+\Delta \epsilon)$ \cite{KIM02b}.

Still neglecting exchange, Sommerfeld derived a closed-form solution for the electron-ion Bremsstrahlung process (inelastic collision of an electron with a Coulomb point charge \cite{SOMMERFELD53,BIEDENHARN56,BETHE57}) and Elwert \cite{BETHE57} proposed a correction to the PWB cross-section  that approximately reproduced the behaviour of Sommerfeld's solution. The latter correction factor is called the Elwert-Sommerfeld factor (denoted ES in the following) and has been successfully used to correct the PWB cross-section for electron-ion Bremsstrahlung of fully ionized atoms. Invoking the similarity between Bremsstrahlung and electron-ion scattering, Jung \cite{JUNG92} suggested to use the ES factor to correct the PWB electron-impact-excitation cross-section. The Elwert-Sommerfeld correction factor $f_{\mathrm{ES}}$ is

\begin{equation*}
f_{\mathrm{ES}}(\epsilon)=\sqrt{\frac{\epsilon}{\epsilon-\Delta \epsilon}}\,\frac{1-e^{-2\pi Z_a/\sqrt{2\epsilon}}}{1-e^{-2\pi Z_b/\sqrt{2(\epsilon-\Delta \epsilon)}}},
\end{equation*}

\noindent where $Z_a$ and $Z_b$ are the initial and final effective charges, respectively. Kilcrease and Brookes \cite{KILCREASE13} suggest to take $Z_a=Z_b=Z$, where $Z$ is the effective ion charge, thus giving

\begin{equation*}
f_{\mathrm{ES}}(\epsilon)=\sqrt{\frac{\epsilon}{\epsilon-\Delta \epsilon}}\times\frac{1-e^{-2\pi Z/\sqrt{2\epsilon}}}{1-e^{-2\pi Z/\sqrt{2(\epsilon-\Delta \epsilon)}}}.
\end{equation*}

\begin{figure*}
\vspace{1cm}
\begin{center}
\includegraphics[width=10cm]{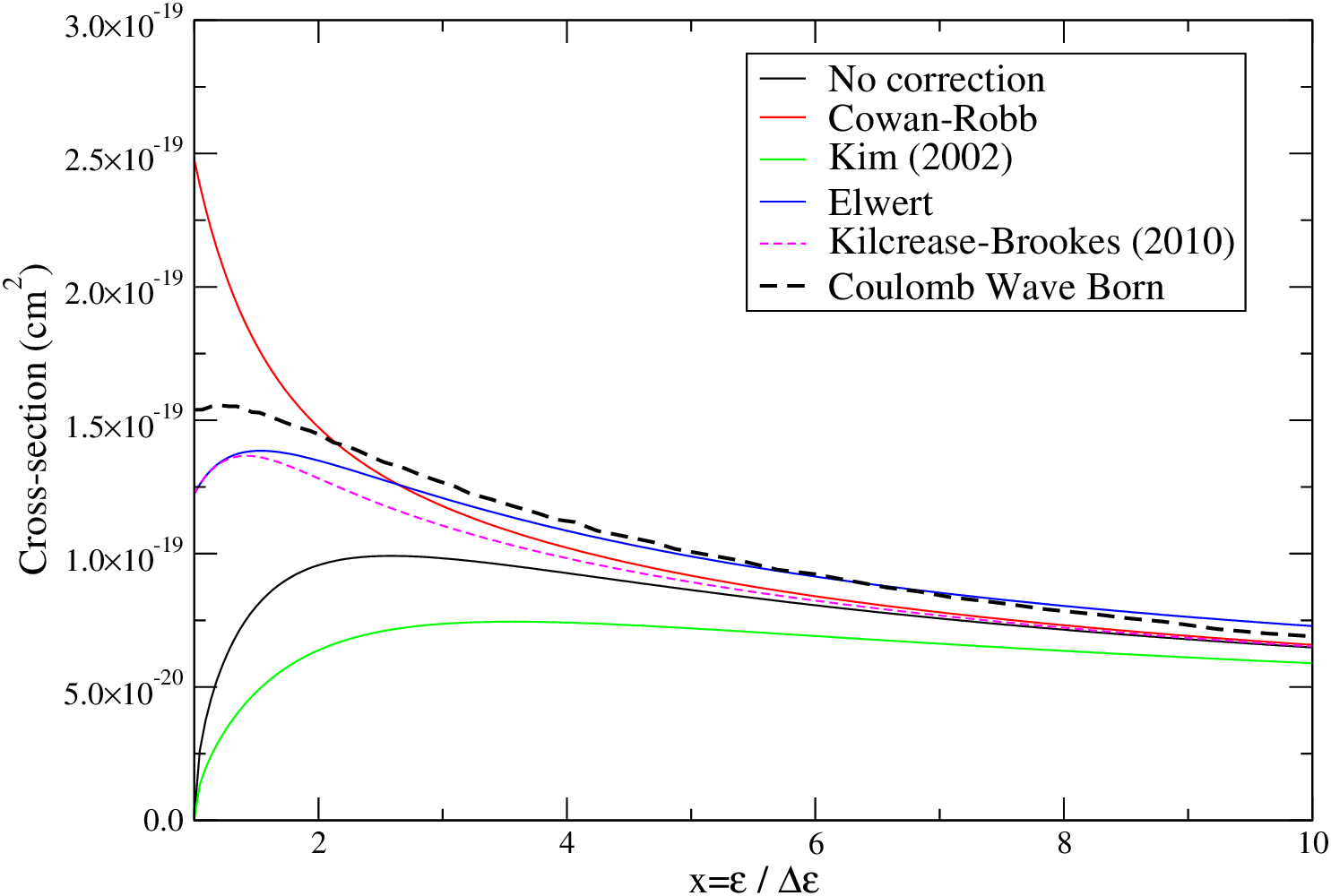}
\end{center}
\caption{(Color online) EIE cross-section with different near-threshold correction factors in [Li] C IV (transition $1s^22s\rightarrow 1s2s2p$). Reference: Coulomb Wave Born \cite{SEATON62}. The transition energy is $\Delta\epsilon=296 eV$.}\label{fig2bis}
\vspace{1cm}
\end{figure*}

\noindent The factor $f_{\mathrm{ES}}$ slowly tends to 1 as $\epsilon$ increases and the latter authors have found that by accelerating this behaviour with the use of the heuristic replacement $Z\rightarrow Z\Delta \epsilon/\epsilon$ improved the agreement with CWB cross-sections at higher $\epsilon$ values. Figure \ref{fig2bis} displays a comparison between the PWB cross-section, the cross-section corrected by different factors: Cowan-Robb \cite{COWAN81}, Kim \cite{KIM02a,KIM02b}, Elwert \cite{SOMMERFELD53,BIEDENHARN56,BETHE57} and Kilcrease-Brookes \cite{KILCREASE13} and a CWB computation \cite{ZHANG89} which is our reference here. In the present case, Elwert's formulation seems to provide the best agreement.

\section{Plasma density effects: shift due to electrons}\label{sec4}
In order to take into account plasma density effects on level energies, several analytical formulas were obtained (see for instance the non-exhaustive list of references \cite{LI12,MASSACRIER90,POIRIER15,BELKHIRI15}) to approach ion-sphere potentials. Using first-order perturbation theory together with hydrogenic scaled mean ionization yields analytical formulas to estimate energy-level shifts.

\subsection{Massacrier-Dubau approach}\label{subsec41}

Assuming a uniform electron gas in the Wigner-Seitz sphere (of radius $R$), Massacrier and Dubau obtained the energy-level shift associated with the subshell $(n\ell)$ \cite{MASSACRIER90}:

\begin{equation}\label{r2}
\Delta\epsilon_{n\ell}=\epsilon_c\left(3-\frac{\langle r^2\rangle_{n\ell}}{R^2}\right),
\end{equation}

\noindent where

\begin{equation}\label{squ}
\langle r^2\rangle_{n\ell}=\frac{n^2}{2Z_{\mathrm{eff}}^2}\left[5n^2+1-3\ell(\ell+1)\right]
\end{equation} 

\noindent and $\epsilon_c=Z^*/(2R)$, $Z^*$ being the average ionization (mean charge) of the plasma. Equation (\ref{r2}) is very simple and easy to handle. For instance, it can be easily implemented in atomic-physics codes \cite{JARRAH17,JARRAH19} or used to estimate the critical electron density at which pressure ionization occurs, \textit{i.e.} at which a bound level disappears into the continuum (see Appendix A).

\begin{table}
\begin{center}
\begin{tabular}{|c|c|c|c|c|c|}\hline
& \;\;MD\;\; & Li 2012 & Li 2012 & Li 2019 \\
& & & modified & $b$=2 modified \\\hline
\;\;$\Delta\epsilon_{1s}$\;\; & 51.1 & 184.5 & 191.6 & 69.2 \\
\;\;$\Delta\epsilon_{2p}$\;\; & 46.7 & 147.5 & 169.5 & 53.6 \\\hline
\end{tabular}
\end{center}
\caption{$\Delta\epsilon_{n\ell}$ (eV) for EIE channel $1s^22s - 1s2s2p$ in Li-like C for different formulations of the energy-level shift.}\label{lilikec}
\end{table}

\subsection{Li-Rosmej approach}\label{subsec42}

In 2012, Li and Rosmej proposed an asymptotic expansion of the potential experienced by an ion subject to free-electron screening in finite-temperature plasmas, in a closed analytical expression \cite{ROSMEJ11} (in atomic units):

\begin{equation}\label{ros}
V_{\mathrm{f}}(r)=4\pi N_e\left\{\frac{R^2}{2}-\frac{r^2}{6}+\frac{4}{3\sqrt{\pi}}\left[\frac{Z^*}{k_BT_e}\right]^{1/2}R^{3/2}-\frac{8}{15\sqrt{\pi}}\left[\frac{Z^*}{k_BT_e}\right]^{1/2}r^{3/2}\right\},
\end{equation}

\noindent where $N_e$ is the free-electron density at the Wigner-Seitz radius $R=\left[3Z^*/\left(4\pi N_e\right)\right]^{1/3}$, $k_B$ the Boltzmann constant and $T_e$ the electron temperature. The energy shift of the $n\ell$ subshell is then obtained by

\begin{equation}\label{delte}
\Delta\epsilon_{n\ell}=\langle n\ell|V_{\mathrm{f}}(r)|n\ell\rangle_{n\ell}=\int_0^{\infty}V_{\mathrm{f}}(r)R_{n\ell}^2\left(r;Z_{\mathrm{eff}}\right)r^2dr,
\end{equation}

\noindent where $R_{n\ell}\left(r;Z_{\mathrm{eff}}\right)$ represents the radial part of the hydrogenic wave-function of the subshell with effective nuclear charge $Z_{\mathrm{eff}}$. The formula involves expectation values of powers of $r$. We use the simplified notation $\langle n\ell|f(r)|n\ell\rangle=\langle f(r)\rangle_{n\ell}$. In 2012, Li and Rosmej, maybe unaware of the fact that an exact formula exists for $\langle r^{3/2}\rangle_{n\ell}$, derived an alternative analytical fit for $V_{\mathrm{f}}(r)$, depending only on $\langle r^2\rangle_{n\ell}$ (Eq. \ref{squ}) and $\langle r\rangle_{n\ell}$:

\begin{equation}\label{radius}
\langle r\rangle_{n\ell}=\frac{1}{2Z_{\mathrm{eff}}}\left[3n^2-\ell(\ell+1)\right].
\end{equation}

\noindent They obtained

\begin{eqnarray}\label{appr}
\Delta\epsilon_{n\ell}&=&\epsilon_c\left\{3-\frac{\langle r^2\rangle_{n\ell}}{R^2}+8\sqrt{\frac{2\epsilon_c}{\pi k_BT_e}}-\frac{16}{5R^{3/2}}\sqrt{\frac{2\epsilon_c}{\pi k_BT_e}}\left[\frac{4}{\sqrt{\pi}}\langle r\rangle_{n\ell}+\frac{1}{10}\langle r^2\rangle_{n\ell}\right]\right\},
\end{eqnarray}

\noindent where $\epsilon_c=Z^*/(2R)$. It was shown very recently \cite{IGLESIAS19a} that the fit given in the Li-Rosmej work was inconsistent with the ion-sphere model, on the contrary to the ``original'' potential in Eq. (\ref{ros}). In addition, the quantity $\langle r^{3/2}\rangle_{n\ell}$ can definitely be expressed analytically. This was also pointed out in Ref. \cite{IGLESIAS19a}, where the author indicates that such a quantity can be obtained following the procedure given in Appendix E of Ref. \cite{SZMYTKOWSKI97} using the generating-function formalism (see for instance the textbook by Bransden and Joachain \cite{BRANSDEN93}) and yielding a complicated expression (in the same paper, a table is provided with particular values displayed in the form of rational fractions). It was recently pointed out that a simple expression for $\langle r^{3/2}\rangle_{n\ell}$ exists \cite{PAIN19}, as a particular case of a relation published by Shertzer in 1991 \cite{SHERTZER91}, who provided an expression for $\langle n\ell|r^{\beta}|n\ell'\rangle_{n\ell}$ for arbitrary $\beta$:

\begin{equation}\label{she}
\langle n\ell|r^{\beta}|n\ell'\rangle_{n\ell}=A_{n,\ell,\ell'}\sum_{i=0}^{n-\ell-1}\frac{(-1)^i\Gamma(\ell+\ell'+3+i+\beta)}{i!(2\ell+1+i)!(n-\ell-i-1)!}\frac{\Gamma(\ell-\ell'+2+i+\beta)}{\Gamma(\ell+3-n+i+\beta)}
\end{equation}

where

\begin{equation*}
A_{n,\ell,\ell'}=\frac{(-1)^{n-\ell'-1}}{2n}\left(\frac{n}{2Z_{\mathrm{eff}}}\right)^{\beta}\left[\frac{(n+\ell)!(n-\ell-1)!}{(n+\ell')!(n-\ell'-1)!}\right]^{1/2},
\end{equation*}

\noindent applying therefore also for off-diagonal terms ($\ell\ne\ell'$). $\Gamma$ is the usual Gamma function. It is worth mentioning that a relativistic equivalent of Eq. (\ref{she}) for $\ell=\ell'$ was published by Salamin in 1995 \cite{SALAMIN95}. In the present case, we have $\ell=\ell'$ and $\beta=3/2$, and we get

\begin{eqnarray}\label{shej}
\langle r^{3/2}\rangle_{n\ell}&=&\frac{(-1)^{n-\ell-1}}{2n}\left(\frac{n}{2Z_{\mathrm{eff}}}\right)^{3/2}\sum_{i=0}^{n-\ell-1}\frac{(-1)^i\Gamma(2\ell+9/2+i)}{i!(2\ell+1+i)!(n-\ell-i-1)!}\frac{\Gamma(7/2+i)}{\Gamma(\ell+9/2-n+i)}.
\end{eqnarray}

Inserting Eqs. (\ref{squ}) and (\ref{shej}) in Eq. (\ref{delte}) gives

\begin{eqnarray}\label{exac}
\Delta\epsilon_{n\ell}&=&\epsilon_c\left\{3-\frac{\langle r^2\rangle_{n\ell}}{R^2}+8\sqrt{\frac{2\epsilon_c}{\pi k_BT_e}}\left[1+\frac{(-1)^{n-\ell}\sqrt{n}}{5\left(2RZ_{\mathrm{eff}}\right)^{3/2}}\sum_{i=0}^{n-\ell-1}\frac{(-1)^i\Gamma(2\ell+9/2+i)}{i!(2\ell+1+i)!(n-\ell-i-1)!}\right.\right.\nonumber\\
& &\left.\left.\times\frac{\Gamma(7/2+i)}{\Gamma(\ell+9/2-n+i)}\right]\vphantom{\sqrt{\frac{Z^*}{\pi Rk_BT_e}}}\right\}.
\end{eqnarray}

\noindent The potential first published by Rosmej \textit{et al.} in 2011 \cite{ROSMEJ11} and which is consistent with the fundamental neutrality requirement of the ion-sphere model as shown by Iglesias \cite{IGLESIAS19a}, can therefore be directly used to derive simple analytical formulas to estimate energy level shifts in dense plasmas. The alternative fit by Li and Rosmej \cite{LI12}, which is not consistent with the ion-sphere model, was motivated by the belief that no analytical expression exists for $\langle r^{3/2}\rangle_{n\ell}$, a statement that was invalidated by Iglesias as well. 

\subsection{Li \textit{et al.} 2019 modified (this work)}\label{subsec43}

In 2019, Li \textit{et al.} \cite{LI19} proposed to use 

\begin{equation}\label{li1924}
\Delta\epsilon_{n\ell}=2\epsilon_c\left\{1+\frac{1}{x-1}-\frac{1}{x-1}\left\langle\frac{r}{R}\right\rangle_{n\ell}^{x-1}\right\},
\end{equation}
\noindent with 
$$ x=3-\frac{b}{\pi}\sqrt{\frac{2\epsilon_c}{k_BT_e}},$$
$$b=2$$
$$\left\langle\frac{r}{R}\right\rangle_{n\ell}^{x-1}=\frac{(-1)^{n-\ell-1}}{2n}\left(\frac{n}{2RZ_{\mathrm{eff}}}\right)^{x-1}~\sum_{i=0}^{n-\ell-1}\frac{(-1)^i}{i!(2\ell+1+i)!}\frac{\Gamma(2\ell+2+i+x)\Gamma(1+i+x)}{(n-\ell-i-1)!\Gamma(\ell+2-n+i+x)}.$$

\begin{figure*}
\vspace{1cm}
\begin{center}
\includegraphics[width=10cm]{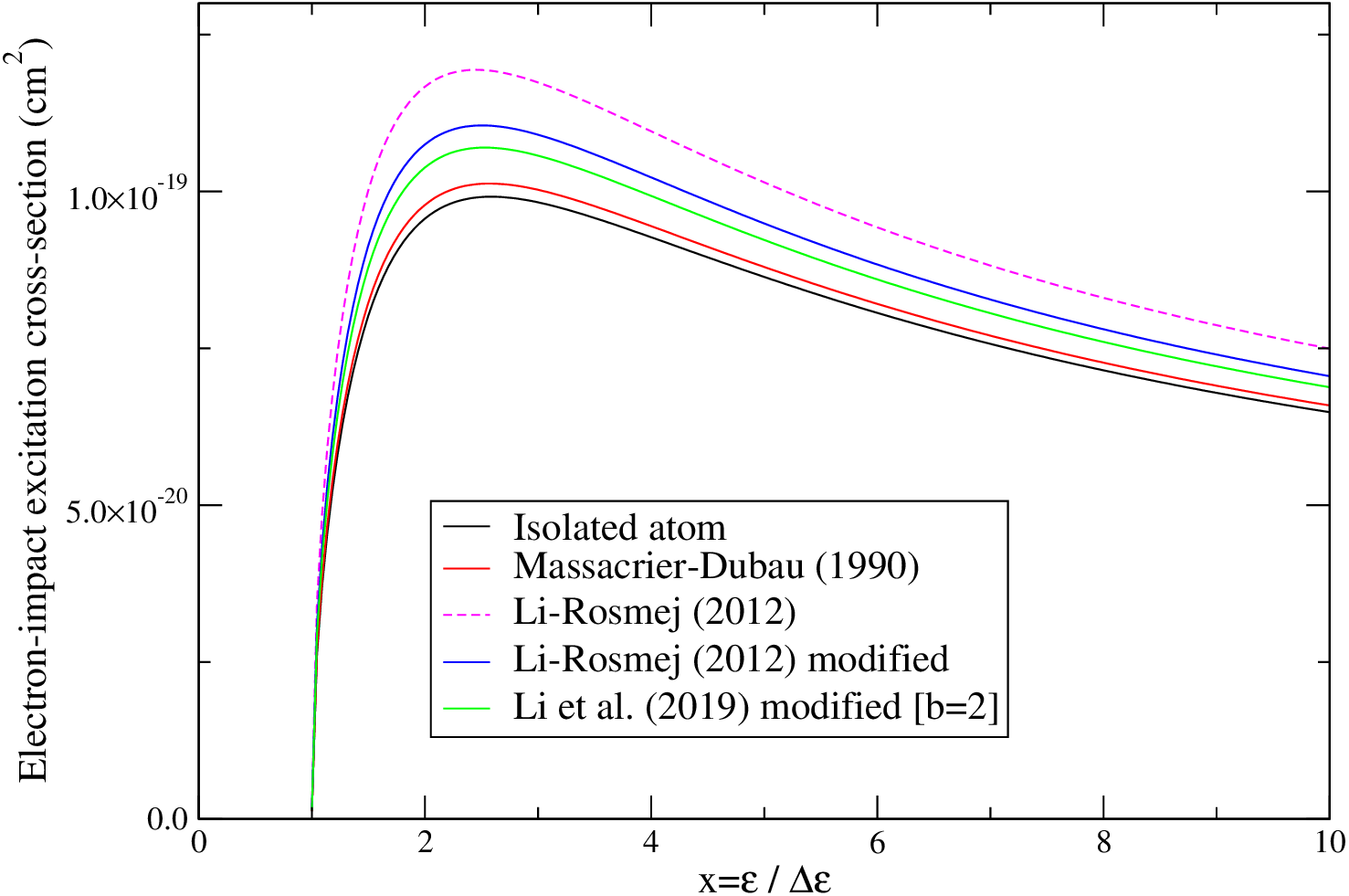}
\end{center}
\caption{(Color online) Effect of different energy-level shifts on EIE cross-section in Li-like C ($1s^22s\rightarrow 1s2s2p$ transition): Massacrier-Dubau (Eq. (\ref{r2})), Li-Rosmej 2012 (Eq. (\ref{appr})), Li-Rosmej 2012 modified (Eq. (\ref{exac})), Li \textit{et al.} modified (Eq. (\ref{li1924})).}\label{fig1bis}
\vspace{1cm}
\end{figure*}

\noindent Figure \ref{fig1bis} shows the effect of the energy-level shift on the EIE cross-section of the $1s^22s\rightarrow 1s2s2p$ transition in Li-like C, for the different formulations mentioned above. The numerical values of the energy-level shift are displayed in table \ref{lilikec}. One can see that, in these conditions, the correction of Massacrier and Dubau is rather small and yields a cross-section which is very close to the isolated-atom one. The difference between the Li-Rosmej (2012) formula and the modified one indicates that the exact computation of the expectation value $\langle r^{3/2}\rangle_{n\ell}$ (see Eq. (\ref{exac})) has a significant effect. It tends to reduce the cross-section, compared to the expression approximated by terms proportional to $\langle r\rangle_{n\ell}$ and $\langle r^2\rangle_{n\ell}$ (see Eq. (\ref{appr}). 

\section{Free-electron degeneracy effects on collisional-excitation rates}\label{sec5}

\subsection{Analytic representation of the cross-section}\label{subsec51}

Many cross-sections were calculated by computer codes or determined experimentally for some values of incident electron energy. Some of them can be found in available atomic databases. However, published cross-sections are often insufficient for detailed simulation of experiments, since data on many cross-sections are missing or do not cover the entire energy range required for calculation of excitation rates, especially for non-Maxwellian plasmas. For this reason, it is desirable to have an easy-to-use formula of known accuracy applicable to various classes of transitions (see for instance the non-exhaustive list of references \cite{VANREGEMORTER62,GOETT80,BUSQUET07,FONTES15}) which is usually represented by the following form of the collision strength \cite{SUNA06}:

\begin{equation}\label{fit}
\Omega\left(\frac{\epsilon}{\Delta \epsilon}\right)=B_0\ln\left(\frac{\epsilon}{\Delta \epsilon}\right)+\sum_{i=1}^5B_i\left(\frac{\epsilon}{\Delta \epsilon}\right)^{-(i-1)}
\end{equation}

\noindent with $i\ge 1$. The logarithm is consistent with the high-energy Bethe limit \cite{BETHE57}. The form of analytical formulas such as Eq. (\ref{fit}) is widely used because it can be analytically integrated over a Maxwellian distribution of electrons. This is important especially for fast codes, which are designed in order to be used ``online'' in radiative-hydrodynamics simulations. Of course, the values of the parameters $B_i$ are specific of a given transition, \emph{i.e.} different transition cross-sections have different fitting parameters. Note that the usual Mewe \cite{mewe72} formulation (see Appendix A) is a particular case of the formalism described here, in the case where the summation of the right-hand side of Eq. \ref{fit} would end at 3 instead of 5. The EIE rate for a non-degenerate electron gas, described by a Maxwell-Boltzmann (MB) distribution, is (see for instance Refs. \cite{ASLANYAN17a,ASLANYAN17b}):

\begin{equation}\label{mb}
\mathcal{R}_{\mathrm{MB}}=\frac{2N_e}{\pi^2}\int_{\Delta \epsilon}^{\infty}\sqrt{\epsilon}~\sigma\left(\epsilon\right)e^{\frac{\mu-\epsilon}{k_BT_e}}d\epsilon,
\end{equation}

\noindent while in the case of a degenerate gas (Fermi-Dirac distribution), it takes the form

\begin{equation}\label{fd}
\mathcal{R}_{\mathrm{FD}}(\Delta \epsilon, T,\mu)=\frac{2N_e}{\pi^2}\int_{\Delta \epsilon}^{\infty}\sqrt{\epsilon}~\sigma\left(\epsilon\right)f(\epsilon,T_e)\left[1-f(\epsilon-\Delta \epsilon,T_e)\right]d\epsilon,
\end{equation}

\noindent where 

\begin{equation*}
f(\epsilon,T_e)=\frac{1}{1+e^{\frac{\epsilon-\mu}{k_BT_e}}}.
\end{equation*}

\noindent The quantity $\left[1-f(\epsilon-\Delta \epsilon,T_e)\right]$ is the Pauli-blocking factor, which takes into account the fact that all final states are not allowed for the free electron. The MB case is recovered if $\mu/(k_BT_e)\ll-1$. Using the relation (\ref{regemo}) between the cross-section $\sigma(\epsilon)$ and the collision strength $\Omega(\epsilon)$ and setting $\eta=\mu/(k_BT_e)<0$ and $\delta=\Delta \epsilon/\left(k_BT_e\right)>0$ in Eqs. (\ref{mb}) and (\ref{fd}), we obtain 

\begin{equation}\label{mbn}
\mathcal{R}_{\mathrm{MB}}=C\int_{1}^{\infty}\Omega\left(x\right)e^{\eta-\delta x}dx
\end{equation}

\noindent and

\begin{equation}\label{fdn}
\mathcal{R}_{\mathrm{FD}}=C\int_{1}^{\infty}\Omega\left(x\right)\frac{1}{1+e^{\delta x-\eta}}\left(1-\frac{e^{\delta}}{1+e^{\delta x-\eta}}\right)dx,
\end{equation}

\noindent where $C$ is a positive constant. Note that the rate of collisional de-excitation from level $i$ to level $j$ would be calculated assuming LTE through detailed balance with collisional excitation

\begin{equation}
\mathcal{R}_{ji}=\mathcal{R}_{ij}\exp\left[\beta\left(\epsilon_j-\epsilon_i\right)\right].
\end{equation}

The purpose of the present study is to compare Eqs. (\ref{mbn}) and (\ref{fdn}). Note that Scott \cite{SCOTT16} and Tallents \cite{TALLENTS16} assume that $\Omega$ is nearly constant, since it is a slowly varying function of energy $E$. In such a way, they estimate the degeneracy effects by defining the ratio

\begin{equation*}
\mathcal{T}(\eta,\delta)=\frac{\int_{1}^{\infty}\frac{1}{1+e^{\delta x-\eta}}\left(1-\frac{e^{\delta}}{1+e^{\delta x-\eta}}\right)dx}{\int_{1}^{\infty}e^{\eta-\delta x}dx}=\frac{e^{-\eta}}{1-e^{-\delta}}\ln\left(\frac{1+e^{\eta}}{1+e^{\eta-\delta}}\right).
\end{equation*}

\begin{table}[ht]
\begin{center}
\begin{tabular}{|c|c|}\hline
Coefficient & Value \\ \hline
$B_0$ & 7.915$\times$10$^ {-3}$ \\ 
$B_1$ & 1.106$\times$10$^ {-3}$ \\
$B_2$ & 2.965$\times$10$^ {-3}$ \\
$B_3$ & 3.247$\times$10$^ {-3}$ \\
$B_4$ & 0 \\
$B_5$ & 0 \\
\hline
\end{tabular}
\end{center}
\caption{Values of the $B_i$ parameters for the $1s-4p$ transition in H-like C.} \label{tab1}
\end{table}

\noindent In the present work, we would like to go one step further, making a less constraining assumption.

\subsection{Maxwell-Boltzmann case}\label{subsec52}

We can write

\begin{equation*}
\mathcal{R}_{\mathrm{MB}}=C_0\mathcal{V}_{0,0}+\sum_{i=1}^5C_i\mathcal{V}_{0,i},
\end{equation*}

\noindent where $C_i=C\times B_i$ for $i\ge 0$.

\begin{equation*}
\mathcal{V}_{0,0}=\int_1^{\infty}\ln x~e^{\eta-\delta x}dx=e^{\eta}\frac{\Gamma(\delta)}{\delta}
\end{equation*}

\begin{equation*}
\mathcal{V}_{0,i}=\int_1^{\infty}\frac{e^{\eta-\delta x}}{x^{i-1}}dx=e^{\eta}\mathrm{E_{i-1}}(\delta)\hspace{1.5cm}{\rm for}\ i>1,
\end{equation*}

\noindent where

\begin{equation*}
\mathrm{E_n}(x)=\int_1^{\infty}\frac{e^{-xt}}{t^n}dt.
\end{equation*}

\noindent We finally obtain the well-known expression used in many collisional-radiative codes:

\begin{equation}\label{rmb}
\mathcal{R}_{\mathrm{MB}}=e^{\eta}\left[C_0\frac{\Gamma(\delta)}{\delta}+\sum_{i=1}^5C_i\mathrm{E_{i-1}}(\delta)\right].
\end{equation}

\subsection{The Fermi-Dirac case}\label{subsec53}

As in the preceding case, we can write

\begin{equation*}
\mathcal{R}_{\mathrm{FD}}=C_0\mathcal{V}_{1,0}+\sum_{i=1}^5C_i\mathcal{V}_{1,i},
\end{equation*}

\noindent with the new integrals

\begin{equation*}
\mathcal{V}_{1,0}=\int_1^{\infty}\frac{\ln(x)}{1+e^{\delta x-\eta}}\left(1-\frac{e^{\delta}}{e^{\delta x-\eta}}\right)dx
\end{equation*}

\begin{equation*}
\mathcal{V}_{1,i}=\int_1^{\infty}\frac{1}{x^{i-1}}\frac{1}{1+e^{\delta x-\eta}}\left(1-\frac{e^{\delta}}{e^{\delta x-\eta}}\right)dx.
\end{equation*}

\noindent In order to calculate $\mathcal{V}_{1,0}$, let us set $z(x)=e^{-\left(\delta x-\eta\right)}$, we then have

\begin{equation*}
\mathcal{V}_{1,0}=\int_1^{\infty}\ln(x)\frac{e^{-\delta}}{\left[1+z(x)\right]\left[e^{-\delta}+z(x)\right]}dx.
\end{equation*}

\noindent Expanding the quantity $\displaystyle e^{-\delta}/\left [ (1+z(x))(e^{-\delta}+z(x))\right ]$ in Taylor series with respect to variable $z(x)<1$, we get

\begin{equation*}
\mathcal{V}_{1,0}=\sum_{p=1}^{\infty}(-1)^{p+1}\left(\frac{1-e^{\delta p}}{1-e^{\delta}}\right)\mathcal{W}(p)
\end{equation*}

\noindent with (see Ref. \cite{GRADSHTEYN80}, p. 573, 4.231-1)

\begin{equation*}
\mathcal{W}(p)=\int_1^{\infty}\ln(x)~e^{-p(\delta x-\eta)}dx=\frac{e^{p\eta}}{p\delta}\Gamma(p\delta).
\end{equation*}

\noindent We follow the same procedure for the determination of $\mathcal{V}_{1,i}$ and get

\begin{equation*}
\mathcal{V}_{1,i}=\int_1^{\infty}\frac{1}{x^{i-1}}\frac{e^{-\delta}}{\left[1+z(x)\right]\left[e^{-\delta}+z(x)\right]}dx,
\end{equation*}

\noindent which is equal to

\begin{equation*}
\mathcal{V}_{1,i}=\sum_{p=1}^{\infty}(-1)^{p+1}\left(\frac{1-e^{\delta p}}{1-e^{\delta}}\right)\mathcal{K}_i(p)
\end{equation*}

\noindent with

\begin{equation*}
\mathcal{K}_i(p)=\int_1^{\infty}\frac{1}{x^{i-1}}~e^{-p(\delta x-\eta)}dx=e^{p\eta}\mathrm{E_{i-1}}(p\delta).
\end{equation*}

\noindent Therefore, the final result for the EIE rate taking into account quantum effects is

\begin{equation}\label{rfd}
\mathcal{R}_{\mathrm{FD}}=\sum_{p=1}^{\infty}(-1)^{p+1}\left(\frac{1-e^{\delta p}}{1-e^{\delta}}\right)e^{p\eta}\left\{C_0\frac{\Gamma(p\delta)}{p\delta}+\sum_{i=1}^5C_i\mathrm{E_{i-1}}(p\delta)\right\}.
\end{equation}

\subsection{Example of application}\label{subsec54}

In the following, we consider an example taken from Ref. \cite{SUNA06} ($1s-4p$ transition in H-like C). The values of the corresponding $B_i$ parameters are listed in table \ref{tab1}. The main issue with this expression (\ref{rfd}) is that it contains an infinite sum. However, it turns out that the truncation of the sum at $p_{\mathrm{max}}=4$ gives a very good precision in many cases relevant for our applications. In many cases, only the first two terms are sufficient to achieve good accuracy, as will be illustrated in the next section. The degeneracy effects can be quantified by the ratio of Eqs. (\ref{rmb}) and (\ref{rfd}):

\begin{eqnarray}\label{bet}
\Lambda(\eta,\delta)&=&\frac{\mathcal{R}_{\mathrm{FD}}}{\mathcal{R}_{\mathrm{MB}}}=\frac{1}{B_0\frac{\Gamma(\delta)}{\delta}+\sum_{i=1}^5B_i\mathrm{E_{i-1}}(\delta)}\nonumber\\
& &\times\sum_{p=1}^{p_{\mathrm{max}}}(-1)^{p+1}\left(\frac{1-e^{\delta p}}{1-e^{\delta}}\right)e^{(p-1)\eta}\left\{B_0\frac{\Gamma(p\delta)}{p\delta}+\sum_{i=1}^5B_i\mathrm{E_{i-1}}(p\delta)\right\},
\end{eqnarray}

\noindent where $p_{\mathrm{max}}$ is the truncation order of the sum. Note that the denominator is in fact the first term of the summation in the numerator. 

\begin{figure*}[ht]
\vspace{1cm}
\begin{center}
\includegraphics[width=8cm]{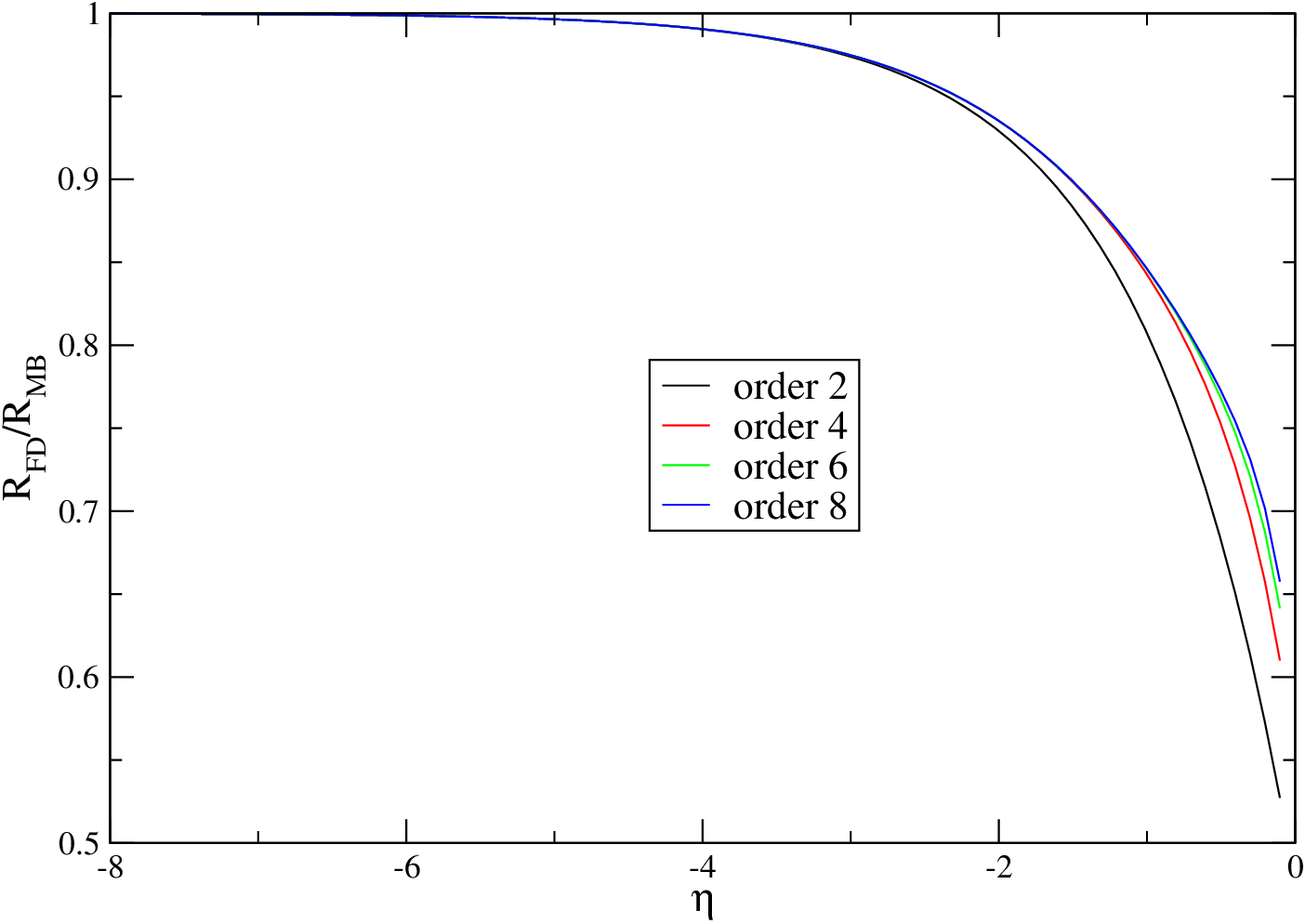}
\end{center}
\caption{(Color online) Values of $\Lambda=\mathcal{R}_{\mathrm{FD}}/\mathcal{R}_{\mathrm{MB}}$ (see Eq. (\ref{bet})) as a function of $\eta$ for $\delta$=0.01 with different truncations of the sum over index $p$: $p_{\mathrm{max}}$=2, 4, 6, 8 and 20.}\label{fig1}
\vspace{1cm}
\end{figure*}

\begin{figure*}[ht]
\vspace{1cm}
\begin{center}
\includegraphics[width=8cm]{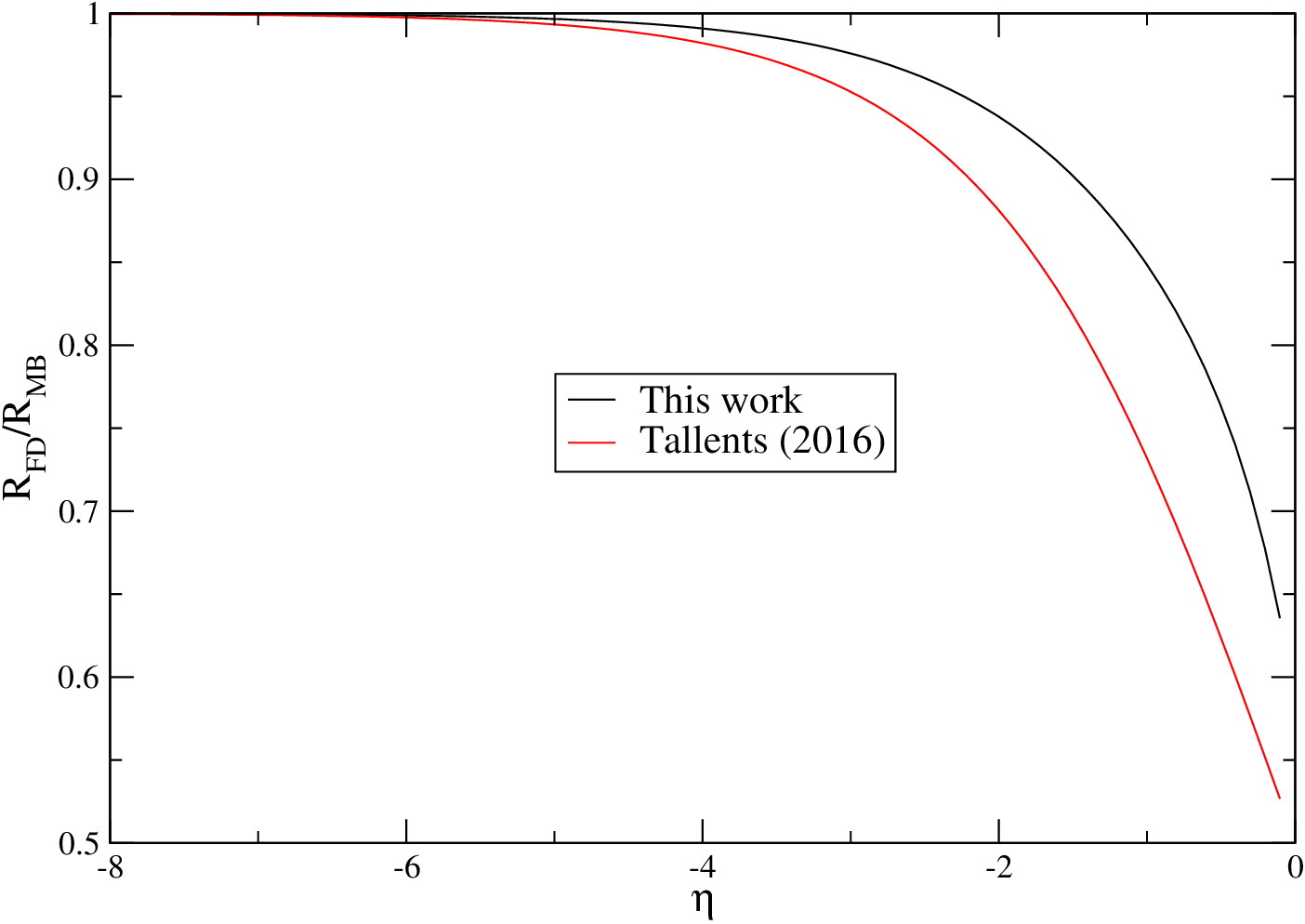}
\end{center}
\caption{(Color online) Values of $\Lambda=\mathcal{R}_{\mathrm{FD}}/\mathcal{R}_{\mathrm{MB}}$ (see Eq. (\ref{bet})) as a function of $\eta$ for $\delta$=0.01. Comparison with the approach of Tallents \cite{TALLENTS16}.}\label{fig2}
\vspace{1cm}
\end{figure*}

\begin{figure*}[ht]
\vspace{1cm}
\begin{center}
\includegraphics[width=8cm]{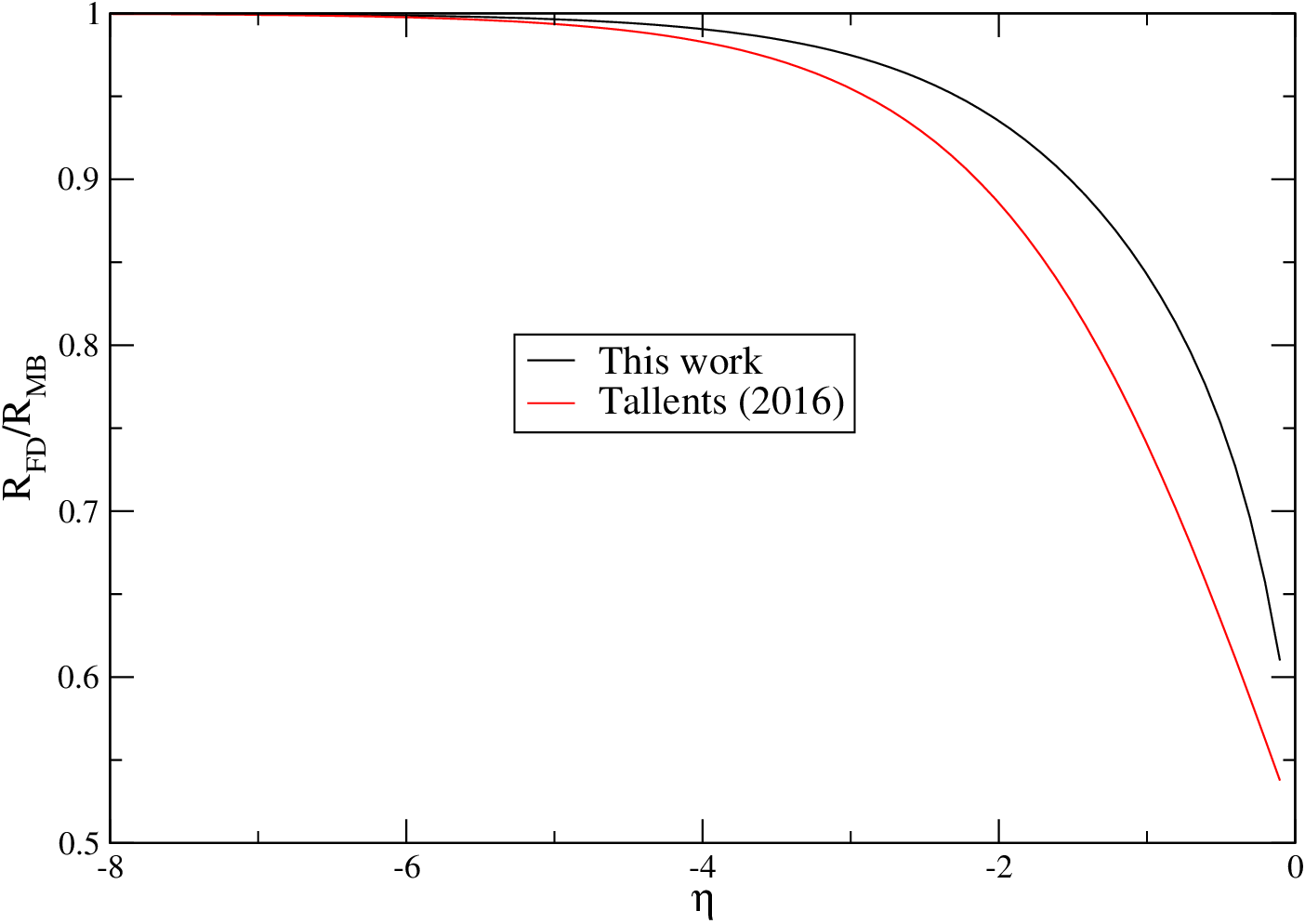}
\end{center}
\caption{(Color online) Values of $\Lambda=\mathcal{R}_{\mathrm{FD}}/\mathcal{R}_{\mathrm{MB}}$ (see Eq. (\ref{bet})) as a function of $\eta$ for $\delta$=0.1. Comparison with the approach of Tallents \cite{TALLENTS16}.}\label{fig3}
\vspace{1cm}
\end{figure*}

\begin{figure*}[ht]
\vspace{1cm}
\begin{center}
\includegraphics[width=8cm]{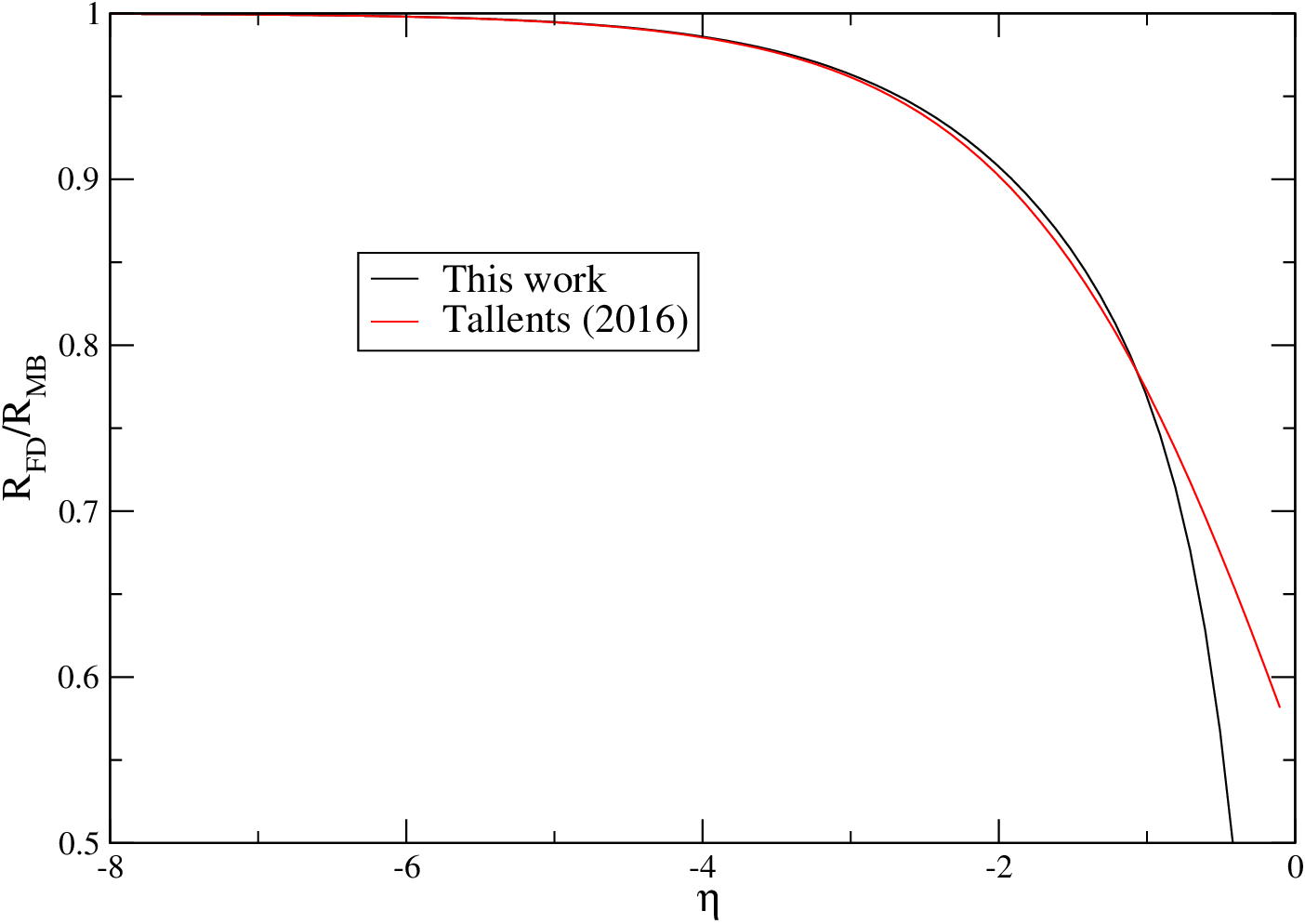}
\end{center}
\caption{(Color online) Values of $\Lambda=\mathcal{R}_{\mathrm{FD}}/\mathcal{R}_{\mathrm{MB}}$ (see Eq. (\ref{bet})) as a function of $\eta$ for $\delta$=0.5. Comparison with the approach of Tallents \cite{TALLENTS16}.}\label{fig4}
\vspace{1cm}
\end{figure*}

\begin{figure*}[ht]
\vspace{1cm}
\begin{center}
\includegraphics[width=8cm]{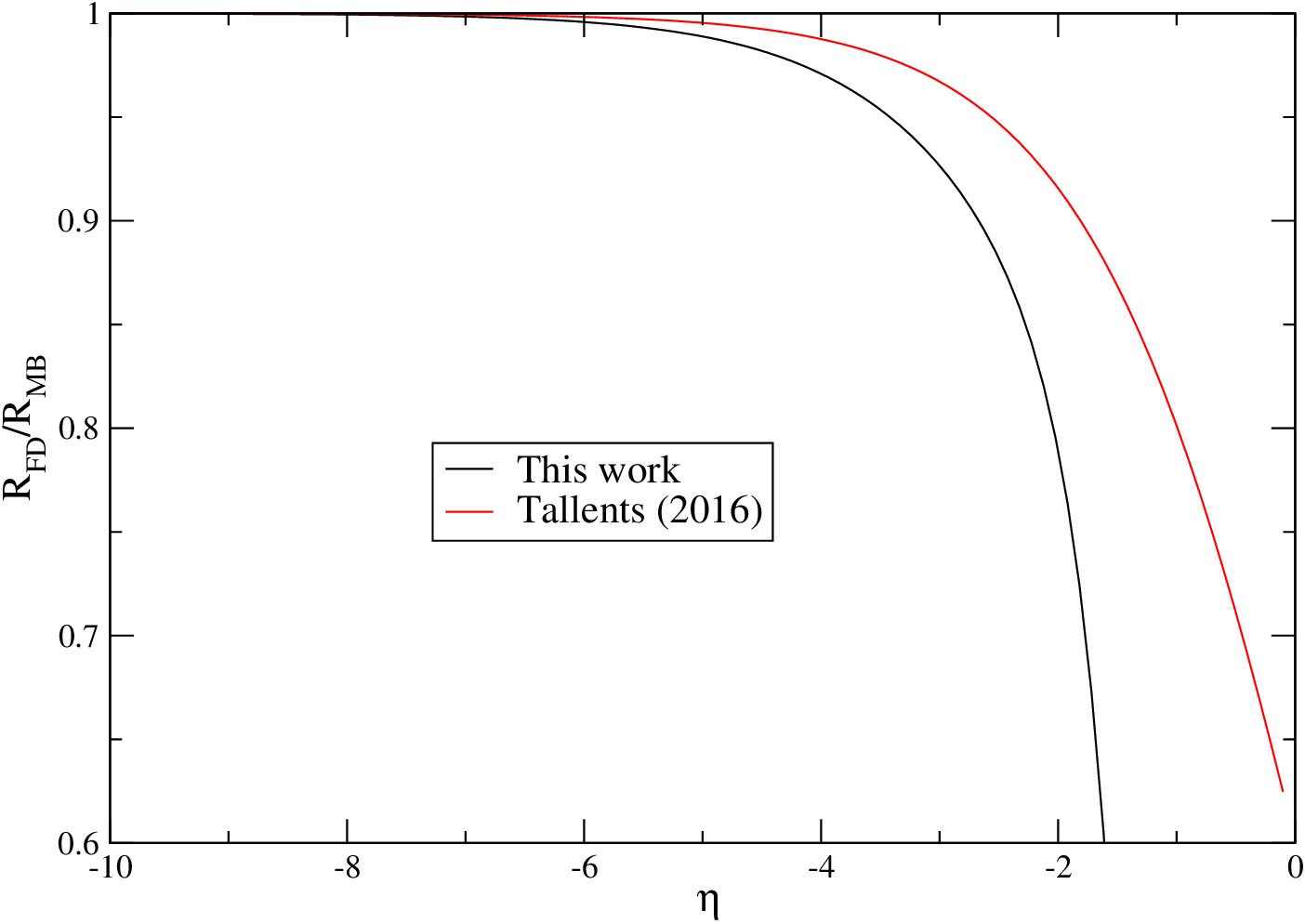}
\end{center}
\caption{(Color online) Values of $\Lambda=\mathcal{R}_{\mathrm{FD}}/\mathcal{R}_{\mathrm{MB}}$ (see Eq. (\ref{bet})) as a function of $\eta$ for $\delta$=1. Comparison with the approach of Tallents \cite{TALLENTS16}.}\label{fig5}
\vspace{1cm}
\end{figure*}

As can be seen in Fig. \ref{fig1} for $\delta=\Delta \epsilon/\left(k_BT_e\right)$=0.01, as well as in tables \ref{tab2} and \ref{tab3} for $\delta$=0.001, 0.01, 0.1 and 0.5, a very good precision is achieved for $p_{\mathrm{max}}$=4. We have checked that this is true over the whole range of relevant values of ($\eta$, $\delta$) in warm dense plasmas.

\begin{table}[ht]
\begin{center}
\begin{tabular}{|c|c|c|c|c|c|}\hline
 $\delta$ / $p_{\mathrm{max}}$ & 2 & 4 & 6 & 8 & 20\\ \hline
0.001 & 0.996631 & 0.996646 & 0.996646 & 0.996646 & 0.996646 \\ 
0.01 & 0.996627 & 0.996642 & 0.996642 & 0.996642 & 0.996642 \\
0.1 & 0.996472 & 0.996489 & 0.996489 & 0.996489 & 0.996489 \\
0.5 & 0.994747 & 0.994787 & 0.994787 & 0.994787 & 0.994787 \\
\hline
\end{tabular}
\end{center}
\caption{Values of $\beta\left(\eta,\delta,p_{\mathrm{max}}\right)$ for $\eta=-5$ and different values of $\delta$ and $p_{\mathrm{max}}$.}\label{tab2}
\end{table}

\begin{table}[ht]
\begin{center}
\begin{tabular}{|c|c|c|c|c|c|}\hline
 $\delta$ / $p_{\mathrm{max}}$ & 2 & 4 & 6 & 8 & 20\\ \hline
0.001 & 0.888427 & 0.902247 & 0.902651 & 0.902665 & 0.902666 \\ 
0.01 & 0.888287 & 0.902141 & 0.902548 & 0.902562 & 0.902563 \\
0.1 & 0.883171 & 0.898313 & 0.898822 & 0.898843 & 0.898844 \\
0.5 & 0.826036 & 0.856105 & 0.859216 & 0.859212 & 0.860068 \\
\hline
\end{tabular}
\end{center}
\caption{Values of $\beta\left(\eta,\delta,p_{\mathrm{max}}\right)$ for $\eta=-1.5$ and different values of $\delta$ and $p_{\mathrm{max}}$.}\label{tab3}
\end{table}

As can be seen in Figs. \ref{fig2}, \ref{fig3}, \ref{fig4} and \ref{fig5}, the new formula (\ref{rfd}) can depart significantly from the one derived by Tallents \cite{TALLENTS16}, especially for $\delta$=1. For small values of $\delta$, the range of reduced chemical potential $\eta$ for which the two approaches differ notably gets smaller and smaller (see Figs. \ref{fig2} ($\delta$=0.01), \ref{fig3} ($\delta$=0.1) and \ref{fig4} ($\delta$=0.5)), but the discrepancy is very important for $\delta$=1 (see Fig. \ref{fig5}).

It is worth mentioning that relativistic effects were studied by Beesley and Rose \cite{BEESLEY19} using the Maxwell-J\"uttner distribution \cite{JUTTNER11,SYNGE}:

\begin{equation*}
F(\epsilon)=\frac{\gamma^2\beta}{\theta K_2(1/\theta)}e^{-\gamma/\theta},
\end{equation*}

\noindent where $\theta=k_BT_e/mc^2$, $\beta=v/c$, $\gamma=1/\sqrt{1-\beta^2}$ and $K_n$ is the modified Bessel function of the second kind:

\begin{equation*}
K_n(z)=\frac{\sqrt{\pi}}{\Gamma(n+1/2)}\left(\frac{z}{2}\right)^n\int_1^{\infty}e^{-zx}\left(x^2-1\right)^{n-1/2}dx.
\end{equation*}

\noindent Such a formula neglects interactions and quantum effects, which is reasonable since relativistic effects become important at high temperature. Beesley and Rose found the correcting factor

\begin{equation*}
\mathcal{R}_{\mathrm{rel}}\left(\theta,\eta_{T_e}\right)=\sqrt{\frac{\pi}{2}}\,\frac{\sqrt{\theta}e^{-1/\theta}}{K_2(1/\theta)}\,\left[1+\frac{1}{2}\left(\theta+\eta_{T_e}\right)\right],
\end{equation*}

\noindent where $\eta_{T_e}=\epsilon/\left(k_BT_e\right)$. Within the Maxwell-J\"uttner distribution, the kinetic energy is given by

\begin{equation}
E=\int_1^{\infty}\gamma f(\gamma)d\gamma=\frac{1}{\theta K_2\left(\frac{1}{\theta}\right)}\int_1^{\infty}\gamma^3\sqrt{1-\frac{1}{\gamma^2}}e^{-\gamma/\theta}d\gamma
\end{equation}

\noindent and one has

\begin{equation}
E=\frac{K_1\left(\frac{1}{\theta}\right)}{K_2\left(\frac{1}{\theta}\right)}+3.
\end{equation}

At temperatures $T\approx mc^2$, pair production will become relevant; while this might not change the velocity distribution, it makes predictions for other quantities based on the velocity distribution and original particle number wrong \cite{singh13}.

\section{Conclusion}\label{sec7}

The description of NLTE plasmas encountered in different fields of high-energy-density science usually involves collisional-radiative models coupled to radiative-hydrodynamics simulations. Therefore, one requires formulas for the cross-sections and rates of the different atomic processes presenting a good compromise between accuracy and computational cost. In this work, we presented an analytical model for the calculation of the EIE cross-section based on hydrogenic formulas. The source code is available upon request. We obtained a complete analytical expression of the generalized oscillator strength and investigated the sensitivity to the near-threshold corrections as well as to the choice of the screened charges. In addition, we studied the impact of different modelings of the plasma density effects based on recently published formulas. Since the measurements of EIE cross-sections in dense plasmas are definitely scarce, it is difficult to provide clear prescriptions, concerning screening charges, near-threshold corrections of electronic level shifts. However the present model enables one to get a qualitative (if not quantitative) insight into the cross-section changes induced by such effects and corrections. We also developed a simple and efficient method to study the impact of degeneracy effects on the EIE rate, following the approach of Tallents \textit{et al.} but removing an approximation (we do not assume that the collision strength can be considered as constant). The main difficulty stems from the integration of the collision strength multiplied by the Fermi-Dirac distribution and the Pauli blocking factor. We found that, using an analytical fit often used in collisional-radiative models, the rate can be calculated accurately without any numerical integration. The ratio between classical and quantum-mechanical EIE rates can be expressed in terms of Gamma and $\mathrm{E_n}$ functions, which are widely used and can be easily computed. In the future, we plan to make comparisons with experimental EIE cross-sections and to investigate the degeneracy effects in the case of very low temperatures, which is important for the start of radiative-hydrodynamics simulations \cite{DESCHAUD14}.

\section*{Acknowledgements}

We would like to thank H\'ector Omar Di Rocco for pointing out typographical errors in the equation preceding Eq. (6) (formula for $gf(k)$) and in Eq. (6) in the original version of the manuscript. Such errors were corrected in the present version.

\appendix

\section{Mewe approximation}\label{meweapp}

Instead of using the approximate formula (\ref{fit}) for the collision strength, another alternative consists in resorting to the Mewe formula \cite{mewe72} of the cross-section (for consistency reasons, we keep, in this Appendix, the units of Ref. \cite{faussurier09}):

\begin{equation}
\sigma(\epsilon)=4\pi a_0^2\frac{2\pi}{\sqrt{3}}\left(\frac{Ryd}{\Delta\epsilon}\right)^2\frac{f_{\mathrm{osc}}}{g}\frac{g(\epsilon/\Delta\epsilon)}{\epsilon/\Delta\epsilon},
\end{equation}

\noindent where $Ryd$ denotes the Rydberg energy, $\Delta\epsilon$ the excitation energy, $g$ the degeneracy of the initial state, $f_{\mathrm{osc}}$ the oscillator strength and

\begin{equation}
g(u)=A+\frac{B}{u}+\frac{C}{u^2}+D \ln(u),
\end{equation}

\noindent with $B=C=0$, $D=0.28$ and $A=0.15$ for $\Delta n\ne 0$ transitions and $0.60$ for $\Delta n=0$ transition \cite{mewe72,faussurier09}. The electron-impact excitation rate reads

\begin{equation}
\mathcal{R}=N_e\int_{\Delta\epsilon}^{\infty}\sigma(\epsilon)v(\epsilon)n(\epsilon),
\end{equation}

\noindent with

\begin{equation}
v(\epsilon)=\sqrt{\frac{2\epsilon}{m}}
\end{equation}

\noindent as well as

\begin{equation}
n(\epsilon)=\frac{2}{\sqrt{\pi}}N_e\frac{\sqrt{\epsilon}}{\left(k_BT_e\right)^{3/2}}e^{-\frac{\epsilon}{k_BT_e}}
\end{equation}

\noindent when the free electrons are assumed to be non-degenerate. In this case, one has \cite{faussurier09}:

\begin{equation}
\mathcal{R}=16 Ryd^2a_0^2c\left(\frac{2\pi^3}{3mc^2}\right)^{1/2}\frac{f_{\mathrm{osc}}}{g}\frac{N_e}{(k_BT_e)^{3/2}}\frac{e^{-\frac{\Delta\epsilon}{k_BT_e}}}{\left[\Delta\epsilon/(k_BT_e)\right]}G\left(\frac{\Delta\epsilon}{k_BT_e}\right),
\end{equation}

\noindent where

\begin{equation}
G(u)=A+(Bu-Cu^2+D)e^uE_1(u)+Cu.
\end{equation}

\noindent For degenerate free electrons, $n(\epsilon)$ becomes

\begin{equation}
n(\epsilon)=\frac{2}{\sqrt{\pi}}N_e\frac{\sqrt{\epsilon}}{\left(k_BT_e\right)^{3/2}}\frac{1}{1+e^{\frac{\epsilon-\mu}{k_BT_e}}},
\end{equation}

\noindent and the rate can be approximated by the same procedure as the one described in section \ref{subsec53}.
Note that, in the present work, we do not use any degeneracy reduction. However, we have the possibility to include it in our code, using the prescription of Zimmerman and More \cite{zimmerman80} consisting in replacing the degeneracy $g_{n\ell}$ of subshell $n\ell$ by

\begin{equation}
\tilde{g}_{n\ell}=\frac{g_{n\ell}}{1+\left(c_1\frac{\langle r\rangle_{n\ell}}{R}\right)^{c_2}},
\end{equation}

\noindent where $c_1$ and $c_2$ are free parameters. The latter parameters are determined by consistency with the zero-temperature Thomas-Fermi model at solid density and at very high density using the numerical fits provided by More \cite{more85}. $R$ is the Wigner-Seitz radius and $\langle r\rangle_{n\ell}$ is given by Eq. (\ref{radius}).

\section{A simple model for pressure ionization}\label{sec6}

Using the simple approach of Massacrier and Dubau (see Sec. \ref{subsec41}), the one-electron hydrogenic energy with plasma effects reads:

\begin{equation*}
\epsilon_{n\ell}=-\frac{Z^2}{2n^2}+\frac{Z}{2R}\left(3-\frac{\langle r^2\rangle_{n\ell}}{R^2}\right),
\end{equation*}

\noindent with $\langle r^2\rangle_{n\ell}$ given by Eq. (\ref{squ}). Using $4\pi R^3N_e/3=Z$ and setting

\begin{equation*}
x=\frac{n^2}{Z}\left(\frac{4\pi N_e}{3Z}\right)^{1/3},
\end{equation*}

\noindent the equation giving the last bound level, $\epsilon_{n\ell}=0$, yields

\begin{equation*}
x^3-3\beta_{n\ell}x+\beta_{n\ell}=0,
\end{equation*}

\noindent with 

\begin{equation*}
\beta_{n\ell}=\frac{n^4}{Z^2\langle r^2\rangle_{n\ell}}=\frac{2n^2}{5n^2+1-3\ell(\ell+1)}.
\end{equation*}

\noindent We use Cardan's method for cubic equations of the type

\begin{equation*}
x^3+px+q=0,
\end{equation*}

\noindent where $p=-3\beta_{n\ell}$, $q=\beta_{n\ell}$. In the case where the discriminant $\Delta=-(4p^3+27q^2)$ is positive, we have three real roots. In the present case $\beta_{n\ell}$ has the minimum

\begin{equation*}
\beta_{\mathrm{min}}=\frac{2n^2}{5n^2+1}.
\end{equation*}

\noindent Therefore the discriminant $\Delta$ reads

\begin{equation*}
4\beta_{\mathrm{min}}^2-1=\frac{3n^2-1}{5n^2+1}
\end{equation*}

\noindent and we are indeed in the case $\Delta\geq 0$. We find it particularly convenient to use the so-called ``trigonometric'' form of the solutions of the cubic equation

\begin{equation*}
x_i=2\sqrt{-\frac{p}{3}}\cos\left[\frac{1}{3}\arccos\left(-\frac{3q}{2p}\sqrt{-\frac{3}{p}}\right)+\frac{2i\pi}{3}\right]
\end{equation*}

\noindent for $i$=0, 1 or 2, \textit{i.e.}

\begin{equation*}
x_i=2\sqrt{\beta}\cos\left[\frac{1}{3}\arccos\left(-\frac{1}{2\sqrt{\beta}}\right)+\frac{2i\pi}{3}\right],
\end{equation*}

\noindent and the critical density of pressure ionization is

\begin{equation}\label{solut}
N_{e,c}=\frac{Z}{n^2}\left(\frac{3Z}{4\pi x_2}\right)^{1/3}.
\end{equation}

\noindent A comparison between the critical density given by formula (\ref{solut}) and the one obtained from a self-consistent-field calculation (SCO-RCG code \cite{PAIN15}) for several subshells in different conditions is provided in table \ref{crit}. Of course, since we assume $Z^*=Z$, the estimate gives poor results when there are few remaining bound electrons, but the formalism presented here may be improved using screened nuclear charges.

\begin{table}[ht]
\begin{center}
\begin{tabular}{|c|c|c|c|c|c|}\hline
Element & $Z^*$ & Subshell & \;\;$T$ (eV)\;\; & \;\;$N_e$, SCO-RCG (cm$^{-3}$)\;\; & $N_e$, Eq. (\ref{solut}) (cm$^{-3}$) \\\hline
Al & 12.9 & $3d$ & 2000 & 2.53$\times 10^{24}$ & 2.87$\times 10^{24}$\\
Fe & 25.7 & $5f$ & 3000 & 2.20$\times 10^{24}$ & 2.23$\times 10^{24}$\\
Fe & 25.95 & $4d$ & 12000 & 9.28$\times 10^{24}$ & 8.77$\times 10^{24}$\\
Fe & 19.5 & $4d$ & 500 & 3.22$\times 10^{24}$ & 8.77$\times 10^{24}$\\\hline
\end{tabular}
\end{center}\caption{Critical electron density of pressure ionization for several subshells in different conditions.}\label{crit}
\end{table}

\clearpage

\end{document}